\documentstyle[art12]{article}

\begin{document}
\pagestyle{empty}

\noindent
ICHEP 94 Ref. gls0656 \hfill OCHA-PP-44

\noindent
Submitted to Pa 12 \hfill June (1994)

\noindent
\ \ hskip 2.4cm Pl 18

\vfill

 \begin{center}
     {\LARGE  Electroweak Baryogenesis and \\
            the Phase Transition Dynamics}\\
      \vfill
      \vspace{0.1in}
    {\large Azusa Yamaguchi and Akio SUGAMOTO}\\
    \vspace{0.1in}
        {\it Department of Physics, Faculty of Science\\
         Ochanomizu University \\
        1-1 Otsuka 2, Bunkyo-ku, Tokyo 112, Japan}\\
        \vspace{0.2in}
 \end{center}
 \vfill
\begin{abstract}
The baryogenesis is reanalyzed based on the model by A.G.Cohen et al., in which
the lepton number, generated by the neutrinos' scattering from the bubble walls
appearing in the development of the electroweak phase transition, is converted
to the baryon number excess through the sphaleron transition. A formula
obtained in this paper on the lepton number production rate is correct for the
both thin and thick walls within the linear approximation.

Investigation on the time-development of the first order phase transition is
simulated, including the temporal change of the wall velocity as well as the
fusion effect of the bubbles. The details of such phase transition dynamics are
found to affect considerably the final value of the baryon number excess.

\end{abstract}

\newpage
\pagestyle{plain}
\setcounter{page}{1}

Baryogenesis in the universe was studied originally in the grand unified
theories in the late 70's~\cite{pre}. After the discovery of the baryon number
and lepton number violation mechanism through the sphaleron
transition~\cite{sph}, we have another possibility of generating baryons at the
electroweak (EW) scale~\cite{EW}.

Among the various model of baryogenesis at the EW scale, we take up the model
by A.G.Cohen et al.~\cite{Cohen}, in which the lepton number, generated by the
neutrino's scattering from the bubble walls appearing in the development of the
EW phase transition, is converted to the baryon number excess through the
sphaleron transition.

The purpose of his paper is to find a formula of the lepton number production
rate which is correct for both thin and thick walls when the vacuum expectation
value of the Higgs scaler changes linearly whitin the wall (which may be called
as the linear approximation).

The other aim of ours is to elucidate the time developement of the phase
transition in detail including the temporal change of the wall velocity as well
as the fusion effects of the nucleated bubbles.

\section{Estimation of the Lepton Number Production Rate from the Bubble Wall}

Our starting {\cal L}agragian is that of Cohen et al.~\cite{Cohen}, namely
\begin{equation}
 {\cal L}=-{\cal L}(standard \ \ model)+\Delta{\cal L} ,
\end{equation}
where
\begin{equation}
  \Delta{\cal L}= {1 \above1pt 2} \overline{N}_L i \gamma^{\mu}
\partial_{\mu}N_{L} - {1 \above1pt 2} \phi \overline{N_L^{\cal C}}
\lambda_{M}^{\dagger}N_{L} - H \overline{\psi_L}\lambda_{D} N^{\cal C}_{L} +
(h.c.) . \label{eq:lagra}
\end{equation}

Here $\psi_L$ and $N_{L}^{\cal C}$ represent left-handed lepton doublets and
right-handed neutrinos of G generations, respectively, $\phi$ stands for the
additional singlet Higgs-scalar, and H is the usual doublet one. The Majorana
and Dirac mass terms of the neutrinos are given by the $G\times G$
mixing-matrices $\lambda^{\dagger}_M$ and $\lambda_D$, respectively. (${\cal
C}$ denotes the charge conjugation as usual.)

Since the electroweak phase-transition is of first order, bubbles of the broken
phase are nucleated within the unbroken vacuum when the phase transition begins
and the system becomes supercooling. These nucleated bubbles grow and fuse with
other bubbles; finally the whole space is filled up with the broken phase.
Therefore we have the thermal non-equlibrium, one of the necessary conditions
of the baryogenesis, during the development of this first order phase
transition.

We have, however, two kinds of Higgs scalars, $H$ and $\phi$.

If they acquire vacuum expectation values at different phases, then we have a
complex configuration of the admixture of various kinds of bubbles. In order to
avoid the complexity, we have assumed that there is only one kind of bubbles.
For this purpose, the position-dependency of the vacuum expectation values
$\langle \phi(x) \rangle$ and $\langle H^0(x) \rangle$ near the interface of
bubbles is asuumed to be identical $\langle \phi(x) \rangle$ $\propto$ $\langle
H^0(x) \rangle$. This rather unrealistic assumption is not too bad, if the
lepton number production rate $f_L$ comes mainly from the lighter components of
neutrinos and not from the heavier ones the latter of which feel the
configuration $\langle \phi(x) \rangle$ sensitively.

The {\cal D}irac equation of neutrinos for a given configuration of $\langle
\phi(x) \rangle$ and $\langle H^0(x) \rangle$ reads from Eq(\ref{eq:lagra})
\begin{equation}
  i (\partial_0-\mbox{\boldmath $\sigma$}\cdot\mbox{\boldmath
$\nabla$})\Psi(x)- i\sigma_2M(x)\Psi(x)^\ast=0  \label{eq:dirac} ,
\end{equation}
where
\begin{equation}
  \Psi(x)=\bigl[ \nu_1, \nu_2, \cdots ,\nu_G, N_1, N_2, \cdots ,N_G \bigr]^{T}
,  \label{eq:psi}
\end{equation}
and $\sigma_i (i=1,2,3)$ are the usual Pauli matrices.

In Eq.(\ref{eq:dirac}) the mass matrix $M(x)$ is given by
\begin{equation}
\  M(x) = \pmatrix{
                   0, & \lambda_D\varphi(x) \cr
            \lambda^{T}_{D}\varphi(x), & \lambda_{M}\varphi(x) \cr
                  }
		  \  \label{eq:mass},
\end{equation}
where $\varphi(x)$ stands for the common configuration of $\langle \phi(x)
\rangle$ and $\langle H^0(x) \rangle$ near the bubble wall, taking 0 outside
the bubble ( unbroken phase), 1 inside the bubble ( broken phase), and the
value in between within the wall of the bubble  (interface of the two phases).
Under the normalization adopted here,
\begin{equation}
 \bigl| (\lambda_D)_{ij} \bigr| \ll \bigl| (\lambda_M)_{kl} \bigr|
\label{eq:lam}
\end{equation}
is understood for the see-saw mechanism to work.

For a sufficiently large bubble of the radius $R \gg T^{-1}$ with the typical
temperature $T \sim O(100GeV)$, we can reduce the three spatial variables to
one, say $z$, representing the radial direction of the bubble.
 The decomposition of $\Psi(z)$ in terms of the spin $S_z$,
\begin{equation}
\   \Psi(z) = \left[
                    \matrix{
                         \psi^{\ast}_4(z)e^{ i E t} \cr
                         \psi_1(z)e^{- i E t} \cr
                         }
                         \right]
                    +
                    \left[
                    \matrix{
                         \psi_3(z)e^{- i E t} \cr
                         \psi^{\ast}_2(z)e^{ i E t} \cr
                         }
                         \right]  \
\end{equation}
indicates that the particle $\psi_1$ with $S_z=-1\big/2$ is coming from the
left, a part of which is reflected to the left as the anti-particle $\psi_4$
with the same spin of $S_z=-1\big/2$; Similarly the anti-particle $\psi_2$ with
$S_z=+1\big/2$ is coming from the left, a part of which is reflected as the
particle $\psi_3$ with $S_z=+1\big/2$.

We consider that the regions $ z<0 $ and $ z>\delta_{w} $ belong to the
unbroken and broken phases, respectively, while the region $ o<z<\delta_{w} $
is occupied by the bubble wall.

The reflection coefficients $R$ and $\bar{R}$ for the particle $(\nu, N)$ and
its anti-particle $(\bar{\nu},\bar{N})$ are defined by
\begin{equation}
 \psi_4(0)= R\psi_1(0) \ \  and \ \ \psi_3(0)=\bar{R}\psi_2(0)
\end{equation}
under the boundary comditions of $\psi_1(\infty)=0$ and $\psi_2(\infty)=0$,
respectively.

In the derivation of $R$ and $\bar{R}$ , we need to estimate the path ordered
integral.
\begin{equation}
I \equiv P exp \left( i \int _{0}^{\delta_{w}} dz A(\varphi(z)) \right)
\end{equation}
with $2G \times 2G$ constant matrix $A$.

We approximate it to
\begin{equation}
I_{A} \cong exp \left( i \int_{0}^{\delta_{w}} dz \varphi(z) A \right)
= e^{{i}\eta A} ,  \label{eq:path}
\end{equation}
where $\eta$ is a parameter representing the shape of the wall $(0<\eta<1)$ ;
\begin{equation}
\eta \equiv \frac{1}{\delta_{w}} \int^{\delta_{w}}_0 dz \varphi(z) \
\end{equation}

It should be noted that the approximation (\ref{eq:path}) is exact when
$\varphi(z)$ changes linearly whitin the wall, $0<z<\delta_{w}$, and for
$\eta=1/2$.
Therefore, the approximation (called linear one) is a rather good
approximation.

Now the reflection coefficiets of particle
$\bigl[(\nu,N)\to(\bar{\nu},\bar{N})\bigr]$ and anti-particle
$\bigl[(\bar{\nu},\bar{N})\to(\nu,N)\bigr]$ read
\begin{equation}
 R \cong - U^{T}DU \ \  and \ \ \bar{R} \cong U^{\dagger}DU^{\ast} ,
\end{equation}
where the diagonal $2G\times 2G$ matrix $D$ in the generation space denotes
\begin{equation}
\   D = D(E) =
      \frac{ s + i \bigl(sE - c\eta M_{b} ^{(diagonal)} \bigr) \times
            P^{-1}_\eta \tan (\delta_{w}P_\eta) }
           { c - i \bigl(cE - s\eta M_{b} ^{(diagonal)} \bigr) \times
            P^{-1}_\eta \tan(\delta_{w}P_\eta) } \   \label{eq:dd} .
\end{equation}

Here the diagonal matrices $c_{\eta}$, $s_{\eta}$ and $P_{\eta}$ are defined as
\begin{equation}
 c_\eta=\cosh \theta_\eta , s_\eta=\sinh \theta_\eta ,
\end{equation}
with
\begin{equation}
  \tanh 2\theta_\eta = \eta \bigl(M_{b} ^{(diagonal)} \big/ E \bigr),
\end{equation}
and
\begin{equation}
   P_\eta = \sqrt{E^2 - (\eta M^{(diagonal)}_b)^2} ,
\end{equation}
where $c=c_1$ and $s=s_1$. The detail of the derivation of (\ref{eq:dd}) will
be written elsewhere~\cite{suga}.

Viewing the expression Eq.(\ref{eq:dd}), we can understand that $D$ is the
reflection coefficient for each mass eigen-states $(\nu,N)$ or
$(\bar{\nu},\bar{N})$, and it is sandwiched by the matrices $U$'s which
transform the weak eigen-states to the mass eigen-states,namely ,
\begin{equation}
UM(z \ \ in \ \ the \ \ broken \ \ phase)U^{T} = M^{diagonal}_b.
\end{equation}

Cooperation of the phase shift including in $D$ and of the complex phases in
the mass matrix (or in $U$) makes the difference between particlex $(\nu, N)$
and anti-particles$(\bar{\nu},\bar{N})$. The number of the comples phases in
$U$ is $G(G-2)+1$, so that the minimum number of the generation is $2$ for our
purpose of the baryogenesis.

We can read easily the following properties from our results Eq.(\ref{eq:dd}):
\def\theenumi{\roman{enumi}}
\begin{enumerate}
\item
$|D(E)|^2 \to {1 \above1pt 4} (m_\nu / E)^{2}$ for $m_\nu / E \to 0$,
 which means that for a massless neutrino the barrier or the wall disappears so
that it perfectly transmit the barrier and does not be reflected.
\item
$|D(E)|^2 \to 1$ for $E \le m_\nu$
, that is, such neutrino is completely reflected by the "potential" barrier.
\item Near the threshold $E = m _\nu + \Delta E (\Delta E > 0) $ , we have
\begin{equation}
 \  |D(E)|^2 \sim 1 -2 \sqrt{ \frac{2\Delta E}{m_\nu} }
 \frac{1 + \eta}{1 + \eta \cos 2\bigl(2\sqrt{1-\eta^2} \delta_{w} m_\nu \bigr)}
 \
\end{equation}
giving the cusp behavior $\sqrt{E-m_\nu}$, and damping rapidly for $ E \ge
m_\nu$. This is the origin of "resonance behavior" in the baryon number
production rate so called by Nelson et al.~\cite{Cohen}.
\item Asymptotically for $E \gg m_\nu $,
\begin{equation}
|D(E)|^2 \sim \left( \frac{m_\nu}{E} \right)^{2}
         \{
          \frac{1}{4} \cos^2 \left( \delta_{w} E \right) +
          \left( \frac{1}{2} - \eta \right)^2 \sin^2 \left(\delta_{w}E \right)
         \},
\end{equation}
where a series of "resonances" appear for $\delta_{w}E \neq 0 $.
\end{enumerate}

In general the phase transition occures at $T = O(100GeV)$, which is
approximately the physical Higgs mass $m_H$, and wall width $\delta_{w}$ is
roughly $m_{H}^{-1}$. Therefore, $\delta_{w}E$, and $\delta_{w}P_{\eta}$ are
$O(1)$ and are not negligible, so that we will be troubled with the summing up
a series of "resonances".

Difference of the reflection rates between the two processes $ \nu_{i} \to
\bar{\nu}_{j} $ and $ \bar{\nu}_{i} \to \nu_{j} $ triggers the lepton number
production
\begin{eqnarray}
 \Delta_{ji} &=& |R_{ji}|^2 - |\bar{R}_{ji}|^2 \nonumber \\
             &=& \sum_{k \neq l} Im(D_{k}D_{l}^{\ast})
                 \times J^{lk}_{ji}  \label{eq:delij} ,
\end{eqnarray}
where
\begin{equation}
J^{lk}_{ji} \equiv Im (U_{kj}U_{ki}U_{lj}^{\ast}U^{\ast}_{li})
\end{equation}
is the so called Jarlskog parameter~\cite{Jar}, typically representing the
magnitude of ${\cal CP}$ violation arising from the mass matrix. The expression
eq.(\ref{eq:delij}) shows that in order for the ${\cal CP}$ violation to
manifest itself, the other dynamical phase, $ Im(D_{k}D^{\ast}_{l}) $ is
required to take part in the problem.

Here we will meet with a difficult problem: The initial states ${i}$ and the
final states ${j}$ are thermally averaged, so that if the thermal distribution
is common for all ${i}$ or for all ${j}$, then we have no lepton number
production, $\sum_{i}\Delta_{ji}=\sum_{j}\Delta_{ji}=0 $. This reflects the
${\cal CPT}$ invariance and the GIM cancellation. The way to overcome this
difficulty has been proposed by Farrar and Shaposhnikov~\cite{Shapo}. They has
considered that the initial and final particles have masses due to the finite
temperature effects that lifts the common distribution for initial and final
particles.

Here we restrict ourselves only to the process in which the neutrinos are
coming from the unbroken phase and are reflected by the bubble wall. Then,
initial and final neutrinos are massless at $T=0$, but aquire the following
finite temperature masses for $\nu_{L}$ and $N_{L}$,
\begin{eqnarray}
M_{\nu_{L}}(T^2) &=& \frac{T^2}{16} \lambda_{D} \lambda^{\dagger}_{D}
\\
M_{N_{L}}(T^2) &=& \frac{T^2}{16} (\lambda^{\dagger}_{M} \lambda_{M})^{T}, \\
\end{eqnarray}
without breaking the chiral invariance~\cite{chic}.

Then, we have obtained the lepton number production rate $f_{L}$ per unit time
generated from the unit area of the bubble wall, under the situation that the
wall is moving toward the incoming neutrinos with the velocity $v_{\omega}$ :
\begin{equation}
f_{L} = \frac{1}{\gamma_\omega} \int
 {d^{3}k^{W}\over (2\pi)^3 }
        \sum_{m n} \bigl(1-f_{n}(E'_{n}) \bigr) \Delta L_{nm} (|k^{W}_{\perp}|)
        \times \frac{k^{W}_{\perp}}{E^{W}_{m}}f_{m}(E_{m}),
\end{equation}
with
\begin{equation}
\Delta L_{nm} = \sum_{i,j} (L_{j} - L_{i})
               \{ |a^{\dagger}_{nj}R_{ji}a_{im}|^2 -
                  |a^{\dagger}_{nj} \bar{R}_{ji}a_{im}|^2 \}
\end{equation}
Here, $L_{i(j)}$ denotes the lepton number of the neutrino $\nu_{i(j)}$ ( or
$N_{i(j)}$ ) in the flavor eigenstates, $m(n)$ represent the mass eigenstates
diagonalized by the matrix $\{ a_{im} \}$, $f(E)$ is the Fermi-Dirac
distribution in the thermal frame, the energy and momentum with the affix $W$
stand for the variables in the wall rest frame, and $\gamma_{\omega} =
(1-v_{\omega}^2/c^2)^{-1/2}$.

Approximately we have the following expression
\begin{eqnarray}
f_{L} & \approx & T^3 \sum_{m,n} \frac{M^{2}_{m}(T) - M^{2}_{n}(T)}{2T^2}
       \int^{\infty}_{0} \frac{d(k^{W}_{\perp}/T)}{(2\pi)^2}
       (k^{W}_{\perp} /T) \Delta L_{nm}(|k^{W}_{\perp}|) \nonumber \\
      & \times &
              \{ \ln \gamma_{\omega} +
              [ \ln(\sqrt{(k^{W}_{\perp})^2+M_{m}^{2}(T)} \big/ T)
             + \gamma_{E} + \ln2 \nonumber \\
      &-& 2 \gamma_{\omega} \sqrt{(k^{W}_{\perp})^2+M_{m}^{2}(T)} \big/ T
               \} \label{eq:flac}
\end{eqnarray}
where $ \gamma _{E}$ is the Euler's constant $(= -0.577 \cdots )$.

Then, $f_{L}$ is the sum of $3$ terms, having different dependence on the wall
velocity $v_{\omega}$ (or $\gamma _{\omega}$ );
\begin{equation}
f_{L} \big/ T^3 \approx (A \ln \gamma_{\omega} + B - C \gamma_{\omega} )\times
J,
\end{equation}
where the constants $A,B$ and $C$ depend on the model and the phase transition
temperature $T$.

As an example, we take the $2$ generation model, where only two lighter
neutrinos $\nu_{1}$ and $\nu_{2}$ are assumed to contribute to the lepton
number generation. Using the parameters of $ m_{1} = M_{1}(T) = 100 GeV$,
$m_{2} = M_{2}(T) = 50 GeV$ and $T=100 GeV$, we have
\begin{eqnarray}
A &=& 0.625 \times 10^{-3} \nonumber \\
B &=& 0.110 \times 10^{-2} \nonumber \\
C &=& 0.185 \times 10^{-2} . \nonumber \\
\end{eqnarray}

The same result is obtained in case of doubling the energy scale
as  $m_{1} = M_{1}(T) = 200 GeV $, $ m_{2} = M_{2}(T) = 100 GeV$
and $ T = 200 GeV$.
Diffficulty existing in the calculation of (\ref{eq:flac}) is the integration
over an infinite tower of small "resonances", so that the obtained values are
preliminary. Using the thin wall approximation, there is no such trouble. Final
lepton number production rate depends on $J$, the magnetude of the ${\cal CP}$
violation.

\section{
Temporal Development of the Phase Transition Dynamics
}

Next theme which we are going to study in this paper is the phase transition
dynamics. During the development of the EW phase transiton, the bubbles of the
broken phase are nucleated at a rate of $I$ per unit time and unit volume. They
grow and fuse with each other. Finally the whole space is covered by the broken
phase. The lepton number production occurs at the interfaces between the broken
and unbroken phases, or the bubble walls. Therefore we need to know the
temporal change of the total area $A$ of these bubble walls, that is $A=A(t)$.

If the wall velocity $v_{\omega}$ is constant at any situation, then the total
number of the lepton number production can be obtained without the knowledge of
$A(t)$;
\begin{equation}
N_{L}= \int v_{\omega}^{-1} f_{L}(v_{\omega}) \cdot v_{\omega} A(t) dt =
v_{\omega}^{-1} f_{L}(v_{\omega}) V_{total},
\end{equation}
giving the lepton number density $n_{L}$ of the universe as
\begin{equation}
n_{L}^{(0)} = v_{\omega}^{-1} f_{L}(v_{\omega}) .
\end{equation}

The wall velocity $v_{\omega}$ is, however, by no means a constant,
but is time-dependent \cite{velocity};
\begin{equation}
v_{\omega}(t) = \frac{dR(t)}{dt} = 2 \Gamma \left(
                           \frac{1}{R_c}-\frac{1}{R(t)} \right)
                                         \label{eq:wallvel}
\end{equation}
where $R(t)$ is the radius of the bubble  nucleated at time $t$, and $R_{c}$ is
the critical radius with which the bubble is nucleated. The $\Gamma^{-1}$ is
the friction coefficient which the global bubble feels when it grows in the
heat bath;
\begin{equation}
\frac{d\phi}{dt} = - \Gamma \frac{\delta F[\phi]}{\delta \phi},
\end{equation}
where $F[\phi]$ is the free energy of the Higgs field $\phi$ making the bubble.
Therefore $ v_{\omega}(t) $ increases exponentially $ exp(2\Gamma T/R_{c}^{2})
$ and approaches to the constant velocity of $v_{\omega}(\infty) =
2\Gamma/R_{c}$. Then the typical length scale and time scale in this problem
are $R_{c}$ and $t_{0} = R_{c}^2 / 2\Gamma$. The $\Gamma$ may be estimated
using the linear response theory, and giving $O(T^{-1}) \sim O(100GeV)^{-1})$.

The other important quantity is the nucleation rate $I$ of the bubbles. We have
the expression~\cite{nucl},
\begin{equation}
I = T^4 \left( \frac{F_{c}(T)}{2\pi T}\right)^{3/2} e^{-F_{c}(T)/T}
\end{equation}
with $F_{c}(T)$ , the free energy of the bubble with the critical radius. Using
the finite temperature effective potential estimated at 1-loop level reads
\cite{nucl}
\begin{equation}
V = \frac{\lambda_{T}}{4}\phi^{4} - ET\phi^3 + D(T^2 - T^2_{0})\phi^2
\end{equation}
with
\begin{eqnarray}
D &=& \frac{1}{4v^2} (2m_{W}^2 + m_{Z}^2 + 2m_{t}^2 ) \\
E &=& \frac{1}{\sqrt{2} \pi v^3}(2m_{W}^3 + 3m_{Z}^3) \\
T_{0} &\sim& \frac{1}{2\sqrt{D}}m_{H}
\end{eqnarray}
and
\begin{equation}
\lambda_{T} \sim \lambda \sim \frac{1}{2}(m_{H}/v)^2.
\end{equation}

For the choice of values $m_{W} = 80GeV, m_{Z} = 90GeV, m_{t} = 150(170)GeV $,
$v = 246GeV$ and the unknown parameter $m_{H}=100GeV$ , we have $D \sim
0.27(0.33)$,$E~\sim~0.05$, $T_{0} \sim 100 GeV $ and $ \lambda \sim 0.08$.

Correspondingly, we have roughly
\begin{eqnarray}
\delta_{\omega} & \sim &  2 \sqrt{\lambda}/ET \sim (9 GeV)^{-1} \\
R_{c} & \sim & 2(DT)^3 / \lambda^{3/2} \cdot \epsilon \\
\end{eqnarray}
with the latent heat
\begin{equation}
\epsilon \sim (T_{c}^2 - T^2 ) \phi^{2}_{c} D
\end{equation}
where $T_{c}$ is the temperature below which the broken phase appears with $
\langle~\phi~\rangle=\phi_{c}$. During the temperature $ T_{c} > T > T_{0}$,
the EW phase transition develops. The free energy of the critical bubble
\begin{equation}
F_{c}(T) \sim \frac{2\sqrt{2} \pi E^5 T^5}{3\lambda^{7/2} D^2 (T_{c}^2 - T^2)}
\end{equation}
gives the nucleation rate $I$ as a function of $ x = T_{c} - T$;
Fig~\ref{nuclrate} depicts the function $I=I(x)$, where the vertical scale is
normalized by $R^{3}_{c}t_{0}$ and $x$ is in the unit of $GeV$.
It is a problem of determining a fixed value of the temperature at which the
first order phase transition develops; the problem should be answered by
coupling the phase transition dynamics with the expansion of the universe. Here
we try two typical value $T_{c1} = 100 GeV $ and $T_{c2} = 200 GeV $ for
$T_{c}$. For these values, we have fixed the temperature at
$ T_{1} = T_{c1} - 0.90 GeV $ and $ T_{2} = T_{c2} - 0.29 GeV$ ,
respectively, during the phase transition.

We performed the computer simulation at $T_{1}$ or $T_{2}$ . In the
simulations, we generated the critical bubbles at the rate of $I(T_{1}$ or
$T_{2})$, these bubbles grow by changing their wall velocities according to
(\ref{eq:wallvel}), and the fusion effect of the bubbles is taken into account.

The results of these simulation are given in Fig~.~\ref{rst2} and \ref{rst3}.
The Fig~\ref{rst2} and \ref{rst3} give the temporal evolution of the area of
the wall $A(t)$ and the volume $V(t)$ of the broken phase, respectively.
The value $T_{1}$ and $T_{2}$ are so chosen that the both
simulations of $A(t)$ and $V(t)$ at $T_{1}$ and $T_{2}$ become
identical.

There is the exactly solvable model of Kolmogorov and Avrami \cite{kolav}, in
which the $v_{\omega}$ and $I$ are kept to be constant. This extremely
attractive theory predicts for $D=3$ as
\begin{equation}
A(t) \big/ V_{total} = \frac{4\pi}{3} v_{\omega}^{3}I_{0}t^{3}
          e ^{-\frac{\pi}{3} v_{\omega}^{3}I_{0}t^4} .  \label{eq:kas}
\end{equation}

Our simulation differs considerably from the simple theory of the phase
transition dynamics (\ref{eq:kas}).
In our simulation, each portion $i$ of the bubble walls gives
different time-development, so that the finally produced lepton
number density $n_{L}$ of our simulation becomes
\begin{equation}
n_{L} = \sum_{i} \int f_{L}(v_{\omega}^{i} (t) A ^{i} (t) dt /
V_{total} .
\end {equation}

Then, we have
\begin{equation}
n_{L}/T^{3} = \cases{
    -0.299 \times 10^{-2} \cdot J & for the case 1 \cr
    -0.303 \times 10^{-2} \cdot J & for the case 2 , \cr
    }
\end{equation}
where the parameters used in the case 1 are $T_{c1}=100 GeV$,
 $T_{1}=T_{c1}-0.90 GeV$, $m_{1}=M_{1}(T)=100 GeV$, $m_{2}=
M_{2}(T)=50 GeV $, and $v_{\omega}( \infty )=0.482 $,
whereas those used in the case 2 are
 $T_{c1}=200 GeV$,
 $T_{2}=T_{c2}-0.29 GeV$ , $m_{1}=M_{1}(T)=200 GeV$ , $m_{2}=
M_{2}(T)=100 GeV$ , and $v_{\omega}( \infty )=0.311 $.
The corresponding lepton number densities $n_{L} ^{0} $ obtained
in the Kolmogorov-Avrami model of the phase transition, read
\begin{equation}
n_{L}^{0} /T^{3} = \cases{
    -0.108 \times 10^{-2} \cdot J & for the case 1 \cr
    -0.209 \times 10^{-2} \cdot J & for the case 2 .\cr
    }
\end{equation}

Therefore, we have
\begin{equation}
\frac{n_{L}}{n_{L} ^{0}} = \cases{
    2.769 & for the case 1 \cr
    1.450 & for the case 2 , \cr
    }
\end{equation}
which is the most important results of our paper; The phase
transition dynamics with and without including the change of the
wall velocity are found to affect considerably the final value
of the lepton numer ( as well as the baryon number ) production.

The detailed analysis will be given elsewhere~\cite{suga}.

\section{
Baryogenesis from Leptogenesis
}

One of the difficulty in considering the baryogenesis at EW scale is the
constraint of $m_{H} \le 45GeV$ which guarantees that the sphareron transiton
is suppressed by the expansion of the universe and that the produced baryon
number may not be washed away by the spharelon.

In our case the sphaleron transition should be rapid enough to make the
chemical equlibrium between the lepton number $L$ and the baryon number $B$.
Thanks to the $B-L$ conservation, we have the non-vanishing equlibrium value
for $B$.
\begin{eqnarray}
\langle B \rangle &=& \frac{1}{2}(1+x)\langle B-L \rangle \nonumber \\
                   &=& -\frac{1}{2}(1+x)\langle produced \ \ L\rangle ,
\end{eqnarray}
where $\langle I_{3} \rangle = \langle Y \rangle = 0 $ is assumed with $x =
O(1) ; x=-18/15$ for the light Higgs doublet $H$ and the light singlet scaler
$\phi$ with three fermions of quarks and leptons and, $x=-1/2$  without this
$\phi$ \cite{Cohen}. Anyway we have $n_{B}$ at the same order of magnetude of
the produced $n_{L}$ during the EW phase transition, without the unrealistic
constraint of $m_{H} \le 45GeV$. This is the good point of this baryogenesis
model induced by the neutrinos. To obtain a realistic value for $n_{L} /
n_{\gamma}$ , we need to have {\cal CP} violation $J$ of $
O(10^{-5} \sim 10^{-7} ) $.

\section*{Acknowledgements}
We give our sincere gratitude to Takao Ohta who has guided us to the
interesting field of the phase transition dynamics, with many fruitful
discussions.

We are greateful to Tetsuyuki Yukawa and Hikaru Kawai for giving us the chance
of using the KEK supercomputer.

One of us (A.Y.) gives thanks to Hatsumi Nakazawa, Mami Suzuki and Aya Ito for
their discussions and encouragements.
The other of us (A.S.) deeply thanks to Yoshio Yamaguchi for his many valuable
questions given to us.

This work is supported in part by the Grant-in-Aid for Scientific Research from
the Ministry of Education, Science and Culture (No.05640336).

\begin{figure}[p]
\vspace*{10cm}
\caption[nucleation rate]{nucleation rate I as a function of $x
= T_{c} - T $ [GeV]. The unit of the vertical axis is $ V = T^4
$.}
\label{nuclrate}
\end{figure}%

\begin{figure}[p]
\vspace*{10cm}
\caption[simulation result s1a]{Temporal development of the total area of the
wall for our
simulation and for the Kolmgorov-Avrami theory

(solid line: our simulation, dotted line: Kolmgorov-Avrami)}
\label{rst2}
\end{figure}%

\begin{figure}[p]
\vspace*{10cm}
\caption[simulation result s1v]{Temporal development of the volume fraction of
the broken phase for our simulation and
for the Kolmgorov-Avrami theory

(solid line: simulation, dotted line: Kolmogorv-Avrami)}
\label{rst3}
\end{figure}%

\end{document}